\documentclass[useAMS,usenatbib]{mn2e}
\usepackage{epsfig,subfigure,amsmath,hyperref}
\usepackage{color}

\definecolor{purple}{rgb}{1,0,1}

\bibpunct[; ]{(}{)}{;}{a}{}{,}

\title{Probabilistic image reconstruction for radio interferometers }

\author[P.M. Sutter et al.]
{
\parbox{\textwidth}{
{P.M. Sutter}$^{1,2,3,4}$ \thanks{Email: sutter@iap.fr},
Benjamin D. Wandelt$^{2,3,4,5}$,
Jason D. McEwen$^{6,7}$,
Emory F. Bunn$^{8}$, 
Ata Karakci$^{9}$, 
Andrei Korotkov$^{9}$, 
Peter Timbie$^{10}$, 
Gregory S. Tucker$^{9}$, and 
Le Zhang$^{10}$
}
\vspace{0.4cm}\\
\parbox[c]{\textwidth}{
$^{1}$ Center for Cosmology and Astro-Particle Physics, Ohio State University, Columbus, OH 43210\\
$^{2}$ UPMC Univ Paris 06, UMR7095, Institut d'Astrophysique de Paris, F-75014, Paris, France \\
$^{3}$ CNRS, UMR7095, Institut d'Astrophysique de Paris, F-75014, Paris, France \\
$^{4}$ Department of Physics, University of Illinois at Urbana-Champaign, Urbana, IL 61801\\
$^{5}$ Department of Astronomy, University of Illinois at Urbana-Champaign, Urbana, IL 61801\\
$^{6}$ Department of Physics and Astronomy, University College London (UCL), London WC1E 6BT, UK\\
$^{7}$ Mullard Space Science Laboratory (MSSL), University College London (UCL), Surrey RH5 6NT, UK\\
$^{8}$ Physics Department, University of Richmond, Richmond, Virginia 23173, USA \\
$^{9}$ Department of Physics, Brown University, 182 Hope Street, Providence, RI 02912, USA \\
$^{10}$ Department of Physics, University of Wisconsin, Madison, WI 53706, USA \\
}}

\begin{document}

\maketitle

\label{firstpage}

\begin{abstract} 
We present a novel, general-purpose method for deconvolving and denoising images from gridded radio interferometric visibilities using Bayesian inference based on a Gaussian process model. 
The method automatically
takes into account incomplete coverage of the $uv$-plane, signal 
mode coupling due to the primary beam, and noise mode coupling due to $uv$ 
sampling.
Our method uses Gibbs sampling to efficiently explore the full posterior distribution of the underlying signal image given the data. We use a  set of widely diverse mock images with a realistic interferometer setup and 
level of noise to assess the method. Compared to results from a 
proxy for point source-based {\tt CLEAN}
method we find that in terms of RMS error and signal-to-noise ratio 
our approach 
performs better than traditional deconvolution techniques, 
regardless of the structure of  the source image in our test suite.
Our implementation  scales as $\mathcal{O}(n_p \log n_p)$,  provides full  
statistical and uncertainty information of the reconstructed image, 
requires no supervision,  and provides a robust, consistent framework for 
incorporating noise and parameter marginalizations and foreground removal. 
\end{abstract}


\section{Introduction}

The next generation of large-scale radio interferometers, 
such as ASKAP~\citep{Johnston2008}, 
MWA~\citep{Lonsdale2009}, 
PAPER~\citep{Parsons2010}, and 
SKA~\citep{Jarvis2007},
promise incredible scientific reward but also incredible 
data analysis challenges. The tremendous volume of data, 
high dynamic range, wide bandwidth,
large amounts of radio interference, 
and significant foregrounds present serious instrumentation
and analysis difficulties~\citep{Bhatnagar2009,Norris2013}.
Image deconvolution --- 
the process of removing the effects of signal 
mode coupling due to the primary beam, noise mode coupling due to 
$uv$ sampling, 
incomplete Fourier mode sampling, and noise ---
is an important first step in scientific analysis from these 
instruments~\citep{ThompsonA.Richard2001}.
The ideal algorithm for performing image deconvolution
 suitable for these upcoming surveys would be:
(1) robust to changes in the input signal and instrument 
  configuration, 
(2) parametrized by as few tunable inputs as possible,
(3) as automated as possible,
(4) driven by the \emph{data} rather than user-selected guesses 
   as to the source signal,
(5) as informative as possible given the difficulty of processing the 
   data even a single time,
(6) scalable to the extreme sizes of future data sets, 
(7) as fast as possible,
and (8) able to easily incorporate modeling of foregrounds or other systematics.
 
Unfortunately, the most commonly used algorithm for interferometric 
image reconstruction, 
{\tt CLEAN}~\citep{Hogbom1974}, satisfies few -- if any ---
of these criteria. Originally developed to remove point source foregrounds 
from a smooth background source image, {\tt CLEAN} works by iteratively 
removing the effects of high-intensity point sources in the Fourier-transformed 
$uv$-plane. While the original algorithm excels in a limited number 
of cases, it is not appropriate for general reconstruction 
problems. Modern implementations of {\tt CLEAN} that appear in standard 
radio astronomy packages (e.g., CASA~\citep{Jaegar2008})
incorporate several advanced features such as simultaneous deconvolution 
at multiple scales~\citep{Cornwell2008, Rau2011} and 
time-varying analysis~\citep{Stewart2011,Rau2012}.

However, even these more advanced versions of {\tt CLEAN} 
require significant fine-tuning and supervision during the deconvolution
process. Users must select thresholds for deciding which pixels 
contain ``point sources'' and the amount by which to remove the 
point source effects in $uv$-space. The optimal choices for these 
thresholds are not known in advance for a given observation.
Users may also select specific regions of the image to apply more 
or fewer {\tt CLEAN}ing iterations. {\tt CLEAN} has no final end state: 
users must decide when an image has been deconvolved ``enough'' 
without introducing unwanted artefacts. Finally, {\tt CLEAN} 
gives no information on the uncertainties in the reconstruction.

Realizing the shortcomings of traditional and more 
sophisticated {\tt CLEAN}-based algorithms,
methods based on a regularized likelihood, such as the maximum entropy method ({\tt MEM};~\citealt{Ables1974,Gull1978,Cornwell1985}) generated significant interest.
{\tt MEM} produces  an image estimate which minimizes the 
difference between the  estimate and the data given the level 
of noise and some chosen metric. While {\tt MEM} requires 
less fine-tuning and supervision than {\tt CLEAN}, the user must still 
choose the metric (usually expressed as an entropy functional), and 
the optimal choice of the metric is not known in 
advance~\citep{Starck2001}. 
The final reconstructed image is thus only optimal in regards to that 
metric. Similarly to {\tt CLEAN}, {\tt MEM} gives no uncertainty information
and in general tends to underestimate the 
source image intensity~\citep{Starck2001,Sutton2006}.

As~\citet{Bhatnagar2004} and~\citet{Puetter2005} pointed out, 
optimal reconstruction techniques must be spatially adaptive 
and operate at multiple scales simultaneously.
Many authors have proposed 
relatively new alternative methods based on
compressed sensing techniques
\citep{wiaux:2009:cs, suksmono:2009, wiaux:2010:csstring, wenger:2010, mcewen:riwfov, li:2011a, carrillo:sara, wolz:spread_spectrum, carrillo:purify},
Bayesian processes~\citep{Ayasso2012},
and separation of smooth and point-like components~\citep{Giovannelli2005}.

In this work, we present a  general purpose, Bayesian reconstruction 
algorithm to infer the source image from realistic interferometric 
radio data. Bayesian analysis starts with building a  
generative probabilistic forward model of the data.  
Given this model choice, the posterior probability density 
function (the \textit{posterior distribution}) quantifies what is 
known about the source image once the data have been 
obtained~\citep{Gelman2004}. 
The purpose of this paper is to demonstrate that a surprisingly simple choice of model for the source image, that of 
an \textit{isotropic Gaussian process}, performs very well in realistically simulated examples for a test suite of widely diverse images.  

The key to practical Bayesian image analysis is to be able to navigate 
efficiently through  the very high dimensional parameter space, since 
every pixel value is an independent parameter. 
For example, in this paper we will explore posterior 
distributions in $10^{4}$-dimensional parameter spaces. 
Gibbs sampling is a powerful technique that has been used successfully 
for sky maps with more than $10^{6}$  pixels in the context of Cosmic Microwave 
Background signal reconstruction and power spectrum 
estimation~\citep{Wandelt2004,JewellLevinAnderson2004,Sutter2012}.
Conceptually, Gibbs sampling iterates between samples of the signal 
and its power spectrum in a way that respects the joint posterior 
distribution of signal and power spectrum given the image. 
This separation allows for significant 
speedups compared to grid based evaluations of the 
posterior: the algorithm scales as $n_p \log n_p$, where $n_p$ 
is the total number of pixels, in the ideal pregridded flat-sky limit discussed 
here. 

Our assumptions for our mock observations
 allow us to pregrid the intensities before the iterative 
solution step.
More general curved-sky analysis would necessarily be more expensive, 
either with gridding-regridding steps during the analysis 
in an AW-projection method or  
with the spherical harmonic transform operation in a full 
sky. This is similar to the computational scalability
 of traditional {\tt CLEAN}. However, {\tt CLEAN} typically completes 
in order $\sim 10$ steps, whereas our method usually requires 
$\sim 100$ iterations. 
As we will see below, the primary computational cost in each 
iteration comes from solving a matrix-vector equation. 
Fortunately, this is a common problem in computational science 
and there are many fast, scalable solutions 
available~\citep{PressWilliamH.1986}.

While the Gibbs sampling framework itself is independent of 
prior (see, for example,~\citet{Sutton2006} for an implementation based on 
fluxon models)
we choose an isotropic Gaussian process prior.
Gaussian processes are surprisingly flexible in describing a 
variety of images~\citep{Mackay2003}. 
Because Gibbs sampling can be 
understood as a non-linear generalization of 
the least-squares optimal signal reconstruction provided 
by the Wiener filter (see Section~\ref{sec:method}) without 
requiring a choice of the signal covariance \emph{a priori},
successive samples are always constrained by the data 
in regions of high signal-to-noise. 
In regions of low signal-to-noise, 
the Gaussian process is the least informative completion to a full 
probabilistic model. In this regime the method
still maps out the signal likelihood taking into account all modeled
signals and imperfections in the data at the two-point correlation level. 

In addition to the speed gains mentioned above,
the method of Gibbs sampling offers several advantages 
over traditional deconvolution techniques:
(1) it explores the full posterior shape, giving \emph{complete} statistical 
information on the resulting image;
(2) it \emph{automatically} takes into account signal 
mode coupling from the primary beam;
(3) the sampled representation produced by the method makes
marginalization \emph{trivial};
(4) it provides optimal reconstruction (i.e., Wiener filtering) \emph{without} 
assuming a signal covariance;
(5) since we have already specified the signal prior, the method has \emph{no} 
tunable parameters and operates completely unsupervised;
and (6) it offers a \emph{sparse} reconstruction 
technique with a full Bayesian motivation.

Section~\ref{sec:sims} below presents our mock observations and interferometer 
setup used to assess our new method.
Next, in Section~\ref{sec:method} we outline the method of Gibbs sampling 
as applied to radio interferometers. 
In Section~\ref{sec:comparison} we compare our method 
to a proxy for traditional {\tt CLEAN} in terms of 
reconstructed image fidelity, 
residuals, and statistics such as RMS error and residual signal-to-noise 
levels.
Finally, we conclude in Section~\ref{sec:conclusion} with a discussion 
of planned extensions and 
the potential role of this method in future observations.

\section{Simulated Observations}
\label{sec:sims}

\subsection{Interferometer Setup}
We model the visibility data $d$ obtained from an interferometric observation as
\begin{equation}
  d = I F A~s + I~n,
  \label{eq:data}
\end{equation} 
where $s$ is a vector containing a discretization of the input sky, $A$ is a
primary beam pattern, $F$ is a Fourier transform operator that converts from
pixel space to the $uv$-plane, $I$ is an interferometer pattern in the
$uv$-plane, and $n$ is a Gaussian realization of the noise. 
We discretize the signal $s$, data $d$, and
noise $n$ with $n_p$ elements.

We generate all input visibilities within a
$20$~degree square patch discretized to $128$ pixels per side, similar 
to the approach of~\citet{Myers2003} and~\citet{Sutter2012}.
While
this size of patch violates the strict flat-sky approximation, it allows us to
explore the validity of our technique at higher resolutions and probe the range
of scales accessible to realistic interferometers.
Because it is easier in interferometry to perform calculations in 
visibility space than in image space~\citep{Baron2012},
we do not establish an observation wavelength 
for these mock observations, since 
our results and conclusions are independent of wavelength.

We model the primary beam pattern $A$ as a Gaussian with peak value of unity
and standard deviation $1.5$ deg.
With these parameters the primary beam decreases to a value
of $10^{-3}$ halfway to the edge of the box. This allows us to include all
Fourier modes up to the Nyquist frequency in our analysis and ensures that the
periodic boundary conditions inherent in the Fourier transform do not cause
unwanted edge-effects. We prevent the primary beam from reaching values below 
$10^{-4}$. This value reflects a balance between faithfully representing 
the suppression of signals far from the point center (so that 
Fourier transforms have correct periodic boundaries) and the need 
to preserve numerical stability in the conjugate-gradient algorithm 
utilized in the method below. 

\begin{figure}
  \centering 
  {\includegraphics[type=png,ext=.png,read=.png,width=\columnwidth]{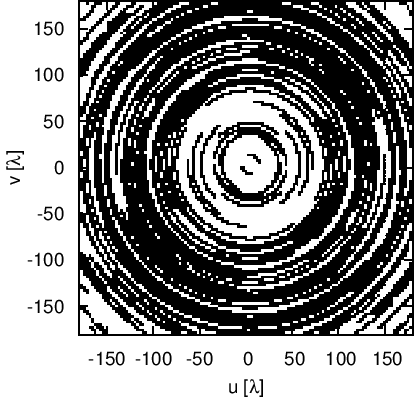}}
  \caption{
           Fiducial interferometer pattern (black) 
           after randomly placing 12 antennas 
           and integrating for six hours assuming a pointing at the 
           north celestial pole.
           The interferometer pattern is discretized to the same 
           resolution as our input images ($n_p = 128$ pixels on a side).
          }  
\label{fig:pattern}
\end{figure}

We assemble the interferometer array in a simple way by randomly placing
12 antennas and selecting all baseline pairs within the $uv$-plane.
We then allow the assembly to rotate uniformly for
6 hours while observing the same sky patch at the north celestial pole.
This choice of antenna arrangement and integration time roughly 
corresponds to existing instruments, such as an extended configuration of
ALMA~\citep{Nyman2010} or VLA~\citep{Perley2011}, although our fiducial 
setup uses fewer elements to highlight the performance of our method 
in less-than-optimal observing regimes.
The interferometer pattern is discretized to the same
resolution as our input images ($n_p = 128$ pixels on a side; described below).
We construct the interferometer
pattern $I$ by placing a value of one wherever a 
baseline length intersects a pixel
during its rotation and zeros elsewhere. We show the resulting $uv$-plane
coverage in Figure~\ref{fig:pattern}.  This configuration covers roughly $70
\%$ of the $uv$-plane, although the coverage varies significantly for each
$\ell$-bin, where $\ell$ is the radius of a given annulus in the 
$uv$-plane. Some bins, especially at
very low and very high $\ell$, have zero coverage due to the lack of baselines
at that distance. However, even if these bins had adequate coverage, we expect
statistics here to be relatively poor due to the reduced number of modes in
these regions. Most bins have at least $60 \%$ coverage and several bins have
complete coverage.

\begin{figure*}
  \centering 
  {\includegraphics[type=png,ext=.png,read=.png,width=0.48\textwidth]{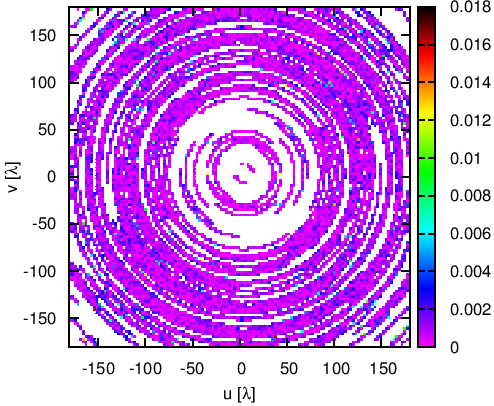}}
  {\includegraphics[type=png,ext=.png,read=.png,width=0.46\textwidth]{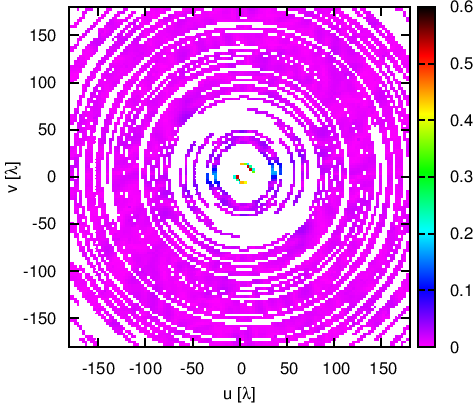}}
  \caption{
           (left) Noise realization used in all test cases. The noise is 
           randomly selected assuming a variance given by
           Eq. (\ref{eq:noise}) and an overall signal-to-noise ratio of 10.
           (right) Example data $d$ (Equation~\ref{eq:data}) in the 
           $uv$-plane for the 
           \emph{Einstein} test image.
          }
\label{fig:noiseanddata}
\end{figure*}

We determine the noise per pixel by summing the integration time spent in that
pixel by all baselines. We do not adopt a noise model for a
particular instrument; rather, we set the noise variance
to be 
\begin{equation}
  \sigma_i^{2} \propto 1/t_{{\rm obs},i},
\label{eq:noise}
\end{equation}
where $t_{\rm obs}$ is the observation time in pixel $i$.
We then set an overall signal-to-noise ratio of $10$
by multiplying all noise variances by a constant value to maintain
$|IFA~s|/|In|=10$.  This provides a scaling of the noise that would
normally be caused by instrument effects such as the 
effective area of the apertures
and the system temperature in a realistic observation.
We use a Gaussian realization of this noise variance to generate
$n$ in Equation (\ref{eq:data}). We then multiply this noise 
by the interferometer pattern $I$ to maintain consistency with the 
signal. 
Figure~\ref{fig:noiseanddata} shows a particular noise realization for 
one of our test images and the resulting data $d$.

\subsection{Test Images}

\begin{figure*}
  \centering 
  {\includegraphics[type=png,ext=.png,read=.png,width=0.24\textwidth]{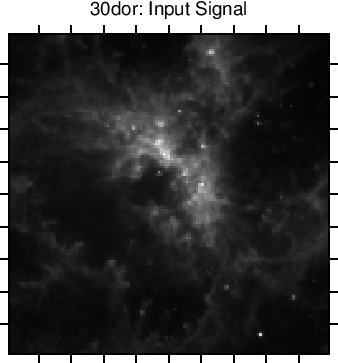}}
  {\includegraphics[type=png,ext=.png,read=.png,width=0.24\textwidth]{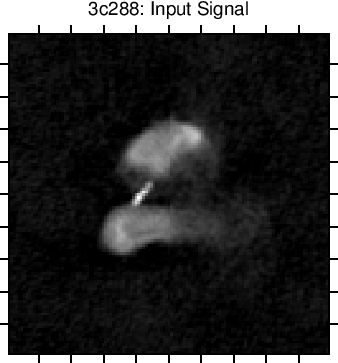}}
  {\includegraphics[type=png,ext=.png,read=.png,width=0.24\textwidth]{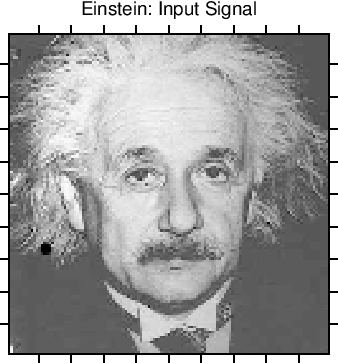}}
  {\includegraphics[type=png,ext=.png,read=.png,width=0.24\textwidth]{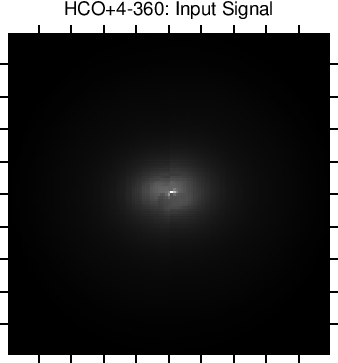}}
  {\includegraphics[type=png,ext=.png,read=.png,width=0.24\textwidth]{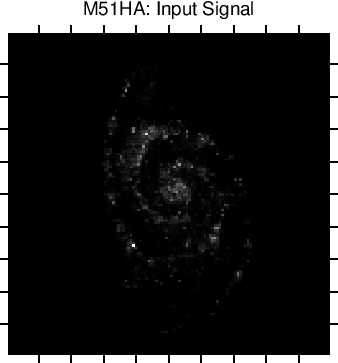}}
  {\includegraphics[type=png,ext=.png,read=.png,width=0.24\textwidth]{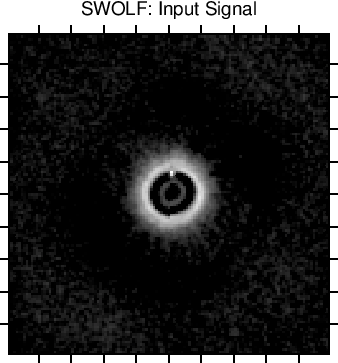}}
  {\includegraphics[type=png,ext=.png,read=.png,width=0.24\textwidth]{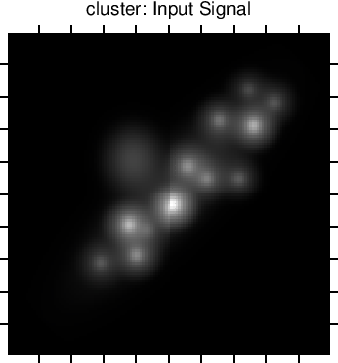}}
  {\includegraphics[type=png,ext=.png,read=.png,width=0.24\textwidth]{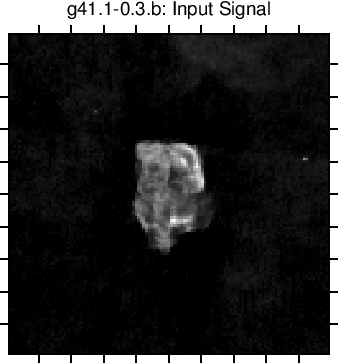}}
  \caption{
           Suite of test input images. All images have been rescaled 
           to $128\times128$ pixels and have had their intensities remapped 
           to a maximum of unity. The repixelization process introduces
           some artefacts, which we leave in place to assess the 
           ability of our isotropic prior to handle anisotropic
           data. Each image is 20 deg across.
            The color scale ranges from $0.0$ (black) to $1.0$ (white).
          }
\label{fig:inputs}
\end{figure*}

Radio interferometers are used to study a wide variety 
of interesting astrophysical and cosmological phenomena, 
such as supernova remnants~\citep{Bhatnagar2011},
molecular gas clouds~\citep{Gratier2010}, 
cluster radio halos~\citep{Cassano2010},
magnetic fields in dwarf galaxies~\citep{Heesen2011},
the interstellar medium~\citep{Zhang2012},
the galactic center~\citep{McClure2012},
and the cosmic microwave background~\citep{PearsonT.J.2000}.
To assess the ability of our method to cope with this variety 
of targets we select eight input images drawn from the 
CASA~\citep{Jaegar2008} user guide
\footnote{\url{http://casaguides.nrao.edu/index.php?title=Sim_Inputs}},
which we present in
Figure~\ref{fig:inputs}.

The test images represent a diverse variety of realistic --- and a few 
unrealistic --- imaging scenarios, such as a protoplanetary disk, 
a face-on spiral galaxy, a cluster, an AGN jet and lobe, and 
the face of Einstein. These images were provided in a 
mix of resolutions and dynamic ranges. To simplify our analysis 
(but without loss of generality), we remapped all images to 
a uniform grid $n_p=128$ pixels on a side and renormalized 
all intensities to a peak of unity.
Some test images had artificial artefacts built-in, and the 
remapping procedure introduced some additional glitches 
in the image. We left these artefacts intact to test the ability 
of our isotropic Gaussian process prior to recover highly 
anisotropic portions of the data.

\section{Method of Gibbs Sampling}
\label{sec:method}

Previous works have extensively discussed Gibbs Samples
~\citep[e.g.,][]{Wandelt2004,Sutter2012}, so we only briefly
introduce the 
relevant equations as applied to interferometric observations here.
We begin with some initial guess of the angular power spectrum $C_\ell^0$ and progressively 
iterate samples from the conditional distributions
\begin{eqnarray}
  s^{i+1} \leftarrow P(s | C_\ell^i, m) \\
  C_\ell^{i+1} \leftarrow P(C_\ell | s^{i+1}),
  \label{eq:iterations}
\end{eqnarray}
where $m$ is the least squares estimate of the signal $s$ given the data $d$ 
(i.e., $B^T N^{-1} B m = B^T N^{-1} d$, where $B$ represents a full 
 signal-to-data operator).
The samples $(C_\ell^i, s^i)$ converge to samples from the 
joint distribution $P(C_\ell,s,m) = P(m|s) P(s|C_\ell) P(C_\ell)$ after a 
sufficient number of iterations.

\begin{figure}
  \centering 
  {\includegraphics[type=png,ext=.png,read=.png,width=0.23\textwidth]{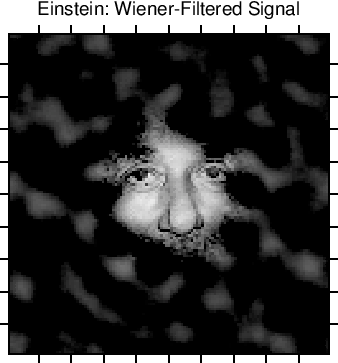}}
  {\includegraphics[type=png,ext=.png,read=.png,width=0.23\textwidth]{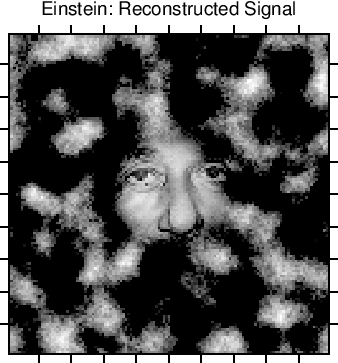}}
  \caption{
           Sample iteration of the Gibbs sampling algorithm for the 
           \emph{Einstein} test image, showing the Wiener-filtered 
           signal (left) and full sample reconstruction with 
           mock fluctuations added (right). Each image is 20 deg across.
           The color scale ranges from $0.0$ (black) to $1.0$ (white).
          }
\label{fig:iteration}
\end{figure}

Given an angular power spectrum sample $C_\ell^i$, we generate a new signal sample 
by drawing from a multivariate Gaussian with mean $S^i(S^i+N)^{-1}m$ 
and variance $((S^i)^{-1} + N^{-1})^{-1}$. 
Here $S$ and $N$ are the signal and noise covariance, respectively.
We do this by solving the set of equations
\begin{equation}
\begin{split}
  M~s^{i+1} = & ~A^T F^{-1} I (INI)^{-1} d  \\
              & + F^{-1} S^{-1/2} F~\xi_1 \\
              & + A^T F^{-1} I (INI)^{-1/2} F~\xi_2,
  \label{eq:sky}
\end{split}
\end{equation}
where we define the matrix operator $M$ as
\begin{equation}
M \equiv F^{-1} S^{-1} F + A^T F^{-1} I (INI)^{-1} IFA.
\end{equation}
In the above equations $A^T$ is the primary 
beam transpose and $F^{-1}$ is the inverse
Fourier transform. The first term in the right-hand side of the above equation
provides the solution for the Wiener-filtered map, while the second and third
terms of Eq.~(\ref{eq:sky}) 
provide random fluctuations with the required variance. The vectors
$\xi_1$ and $\xi_2$ are of length $n_p$ with elements drawn from a standard
normal distribution.
As an illustration, 
Figure~\ref{fig:iteration} shows an example Wiener-filtered map 
(i.e., from just solving the first term in the right-hand side) 
and full signal sample $s^i$ at a single iteration 
for the \emph{Einstein} test case.
We see that the fluctuation terms in Equation~(\ref{eq:sky}) 
fill in regions weakly constrained by the data with a guess that mimics 
the known portions of the signal given the level of noise.

The signal covariance matrix $S$ is diagonal in the $uv$-plane for isotropic
signals (which we assume as part of our Gaussian process prior), 
so $S_{\ell,\ell'} = C_\ell \delta_{\ell,\ell'}$, where $\ell = 2 \pi
u$, with $u$ being the radial distance in the $uv$-plane. Here and
throughout we assume the flat-sky approximation that makes this identity
valid. 
By construction, Gibbs sampling explores the exact posterior and therefore 
treats the couplings  introduced by partial sky coverage 
optimally~\citep{Wandelt2004}. The algorithm ``knows'' about the 
couplings since they are contained in the data 
model (Eq.~\ref{eq:data}) which underlies the analysis.
However, to a very good approximation the noise covariance matrix $N$ is
diagonal, and thus we will assume this for 
simplicity. We assign to the matrix $N$ entries equal to $N_{i,j} = \sigma_i^2
\delta_{i,j}$, where $\sigma_i$ is the noise variance for the $i$th pixel in
the $uv$-plane. The construction $I (INI)^{-1}$ provides a pseudo-inverse 
 of $N$, so
that any locations in the $uv$-plane with no antenna coverage do not yield
infinities when taking the inverse.  

We solve numerically the above matrix-vector equation using a preconditioned
conjugate-gradient scheme~\citep{PressWilliamH.1986}. The preconditioner
approximates the diagonal components of $M$ and is 
\begin{equation}  
P^{-1} = F^{-1} I (INI)^{-1} I F~(F^{-1} \tilde{A}^2), 
\end{equation} 
where $\tilde{A}$ is the Fourier transform of the primary beam pattern.
We implemented the code to solve the above equations with the 
 open-source PETSc library
~\citep{petsc-efficient,petsc-user-ref,petsc-web-page} and the MPI-parallelized
version of FFTW~\citep{Frigo2005}.

Given the latest signal sample, $s^i$, we generate a new angular power spectrum sample
from Eq.~(\ref{eq:iterations}) by computing the variance $\pi_l^2$ in annuli of
constant $\ell$ on the Fourier-transformed signal.  We then use this variance
to draw from the probability density $P(C_\ell | s^i)$, which follows an
inverse Gamma distribution, by creating a vector $p_\ell$ of length $n_\ell$
(assuming a Jeffreys'  ignorance prior) and unit Gaussian random elements.
Here, $n_\ell$ is the number of pixels in the bin $\ell$. The next power
spectrum sample is then simply \begin{equation} C_\ell^{i+1} =
\frac{\pi_\ell}{\left| p_\ell \right|^2}.  \label{eq:clsample} \end{equation}

In the above, we assume $\ell(\ell+1)C_{\ell}$ to be constant across the width
of each annulus in the $uv$-plane. The width of each annulus can be set as
desired. For the test cases which we work with in this paper 
we chose the width to be $8
\pi/L$, where $L$ is the longest baseline of the interferometer. This is four
times the $uv$-space resolution. This choice limits
correlations between angular power spectrum bins which develop as a consequence of
partial sky coverage. All $\ell$-bins have uniform width except for the first,
which we restrict to cover only the central zone where we enforce $C_0^i = 0$, 
since our analysis cannot constrain the DC mode. We wish to capture 
as much power spectrum information as possible, 
so we correspondingly widen the width of the second bin to close this gap.

To determine convergence so that our iterative samples from the conditional
distributions (Eq.~\ref{eq:iterations}) are indeed samples from the joint
distribution, we employ multiple chains with different random number seeds. Our
convergence criterion is the Gelman-Rubin (G-R) statistic, which compares the
variance among chains to the variance within each chain. 
The G-R statistic asymptotes to unity, so convergence is said to
be achieved when this statistic is below a given tolerance level for each
$\ell$-bin~\citep{Gelman1992}.
For our test images we stopped iterating when the G-R statistic 
reached less than $1.1$, which took from 500 to 1500 steps, 
depending on the image. 

\begin{figure}
  \centering 
  {\includegraphics[type=png,ext=.png,read=.png,width=0.28\textwidth]{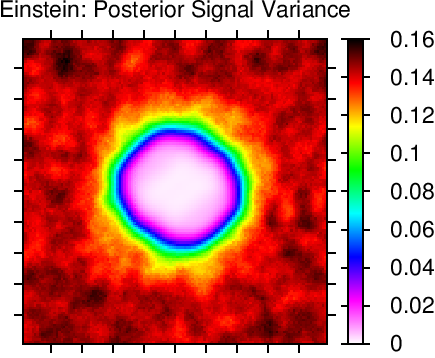}}
  {\includegraphics[type=png,ext=.png,read=.png,width=0.19\textwidth]{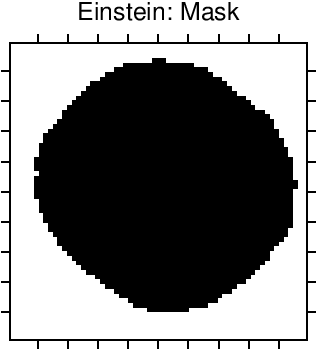}}
  \caption{
           Posterior signal variance in image space (left) and 
           acceptance mask
           derived from the variance (right) for the \emph{Einstein}
           test image. 
           The variance map is 20 deg across, and to 
           show the detailed shape of the variance mask, 
           we have zoomed in on the inner ten degrees.
           The acceptance region of the mask is shown in black.
          }
\label{fig:variance}
\end{figure}

After convergence, we take the mean of the signal samples as 
our reconstructed image and simply calculate the variance 
for each pixel to assess the uncertainty in the reconstruction.
Figure~\ref{fig:variance} shows the variance for the 
\emph{Einstein} test image. We naturally see insignificant variance 
in the center of the primary beam where the data support is strongest, 
with steadily increasing variance away from the pointing center.

We may use the variance to build a mask of the final reconstructed 
image in which we only accept pixels below a given variance threshold. 
While this certainly isn't necessary for the method to work, 
it provides an easy way to combine the posterior mean and 
variance information in a single plot. First, we compute the 
asymptotic variance $\sigma_{\rm asmp}$, 
or the mean of the variance along the outer 
edge of image plane. We then reject any pixel whose variance 
$\sigma$ is greater than $\sigma_{\rm asmp}/2$ \emph{and} 
whose signal-to-noise ratio $s / \sqrt{\sigma} < 1$. This 
last criterion prevents us from masking low-intensity, 
but still well-constrained, pixels. The values chosen to 
create the mask are arbitrary, and we choose them solely 
on aesthetic merit. We repeat: this mask is not necessary to perform 
the deconvolution and produce a reconstructed signal. We only 
use it to neatly incorporate the measured uncertainty information
in the plotted images.

As we see in Figure~\ref{fig:variance}, 
the acceptance mask generally follows the primary beam 
pattern. The detailed shape near the mask edge is influenced 
by the relative signal-to-noise levels and 
highlights the non-obvious nature of the regions of 
reliable reconstruction. We also notice an asymmetry in the 
acceptance mask: this is in response to the asymmetric nature of the 
fiducial interferometer setup (Figure~\ref{fig:pattern}) and provides 
strong evidence that the symmetric nature of our Gaussian 
process prior does not greatly influence our inference.

\section{Comparison to {\tt CLEAN}}
\label{sec:comparison}

We compare images recovered by Gibbs sampling, as described in the previous 
section, to those recovered by a proxy for the point source-based 
{\tt CLEAN} algorithm.  
{\tt CLEAN} is implemented in various radio interferometric imaging 
packages (such as CASA).  However, it is not straightforward to use 
these packages to reconstruct images from simulated visibilities 
already defined on gridded coordinates; these packages are instead 
tailored to analyse observations made by real interferometric 
telescopes, with data in a specific format.  Consequently, 
we compare to reconstructions made with a proxy for 
the point source-based {\tt CLEAN} algorithm.
While there are more sophisticated implementations of {\tt CLEAN} 
that would undoubtedly perform better with our test images, 
this gives us a simple standard of comparison allowing us to 
demonstrate the viability of our method. We will include further comparisons 
in future work.

It was shown by \citet{wiaux:2009:cs} that $\ell_1$ reconstruction with the Dirac basis (i.e.\ pixel basis) results in very similar reconstruction quality to {\tt CLEAN} \citep[see][Figure~1]{wiaux:2009:cs}.  This is to be expected since it is known already that {\tt CLEAN} is closely related to $\ell_1$ reconstruction with the Dirac basis \citep{marsh:1987}.  We thus take a similar approach to that taken by 
recent studies 
(McEwen \& Wiaux 2011; Carrillo et al. 2012; Wolz et al. 2013; Carrillo et al. 2013)
and use $\ell_1$ reconstruction with a Dirac basis as a proxy for the {\tt CLEAN} algorithm.  The reconstructed image that serves as a {\tt CLEAN} proxy is therefore given by the solution of the optimisation problem:
\begin{equation}
  \min \| s \|_1 \ \mbox{such that} \ \| d - IFAs\|_2 \leq \epsilon \ ,
\end{equation}
where $\| \cdot \|_1$ denotes the $\ell_1$ norm and $\epsilon$ is 
related to a residual noise level 
estimator~\citep[see, e.g.,][]{wiaux:2009:cs,carrillo:sara}.  
We solve this problem using the Sparse OPTimisation 
(SOPT\footnote{\url{http://basp-group.github.io/sopt/}}) 
package~\citep{carrillo:sara, carrillo:saraalgo} 
using the Douglas-Rachford splitting algorithm \citep{combettes:2007}.  Note that the SOPT package is a versatile code, capable of solving much more sophisticated optimisation problems than the straight-forward $\ell_1$ minimisation performed here (for examples of more extensive use see 
Carrillo et al. 2012, 2013b; Wolz et al. 2013).

\begin{figure*}
  \centering 
  {\includegraphics[type=png,ext=.png,read=.png,width=0.29\textwidth]{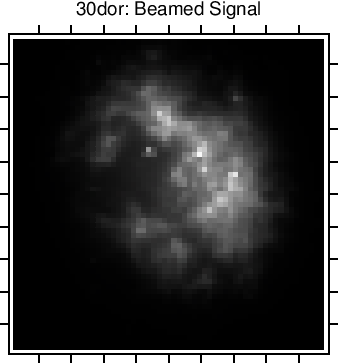}}
  {\includegraphics[type=png,ext=.png,read=.png,width=0.29\textwidth]{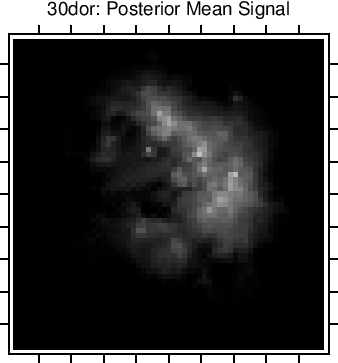}}
  {\includegraphics[type=png,ext=.png,read=.png,width=0.29\textwidth]{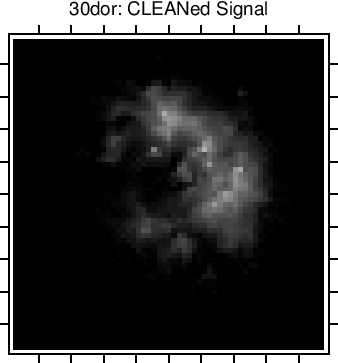}}
  
  {\includegraphics[type=png,ext=.png,read=.png,width=0.29\textwidth]{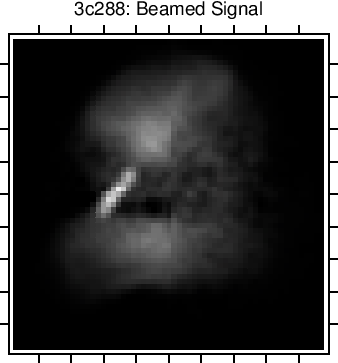}}
  {\includegraphics[type=png,ext=.png,read=.png,width=0.29\textwidth]{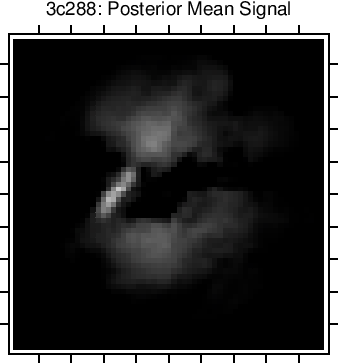}}
  {\includegraphics[type=png,ext=.png,read=.png,width=0.29\textwidth]{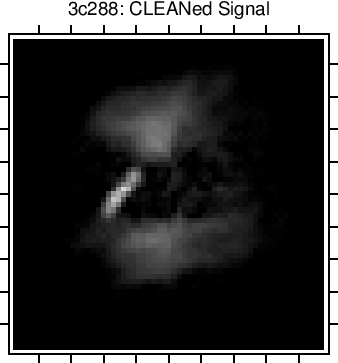}}
  
  {\includegraphics[type=png,ext=.png,read=.png,width=0.29\textwidth]{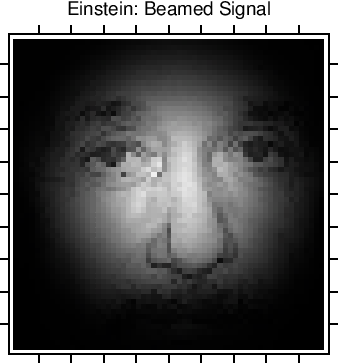}}
  {\includegraphics[type=png,ext=.png,read=.png,width=0.29\textwidth]{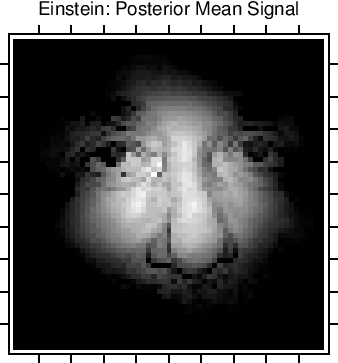}}
  {\includegraphics[type=png,ext=.png,read=.png,width=0.29\textwidth]{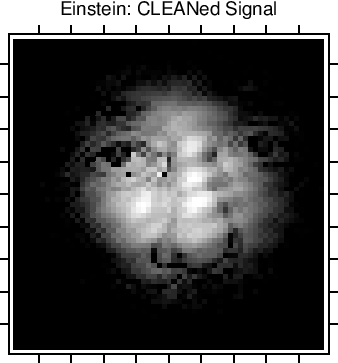}}
  
  {\includegraphics[type=png,ext=.png,read=.png,width=0.29\textwidth]{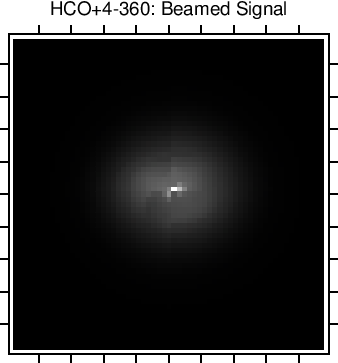}}
  {\includegraphics[type=png,ext=.png,read=.png,width=0.29\textwidth]{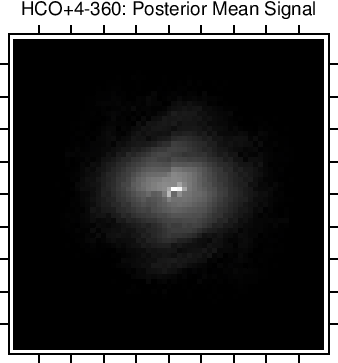}}
  {\includegraphics[type=png,ext=.png,read=.png,width=0.29\textwidth]{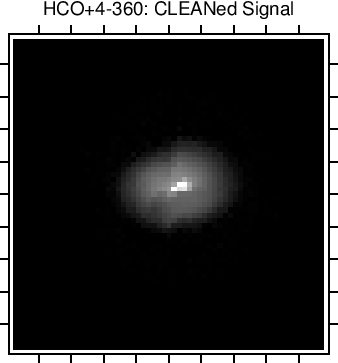}}
  \caption{
           Beamed input images 
           without noise added ($As$; left column), posterior mean signal
           after 500 Gibbs sampling iterations (middle column), and 
           {\tt CLEAN} reconstruction (right column). Axes are marked in deg. 
           For clarity we have zoomed
           in on the innermost 10 degrees where the signal is most 
           prominent.
           The color scale ranges from $0.0$ (black) to $0.6$ (white).
          } 
\label{fig:gallery1}
\end{figure*}

\begin{figure*}
  {\includegraphics[type=png,ext=.png,read=.png,width=0.29\textwidth]{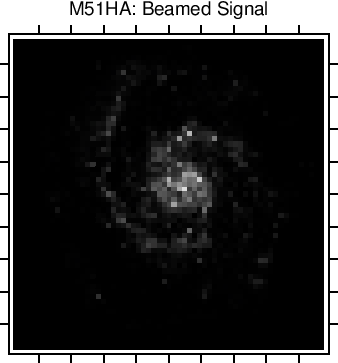}}
  {\includegraphics[type=png,ext=.png,read=.png,width=0.29\textwidth]{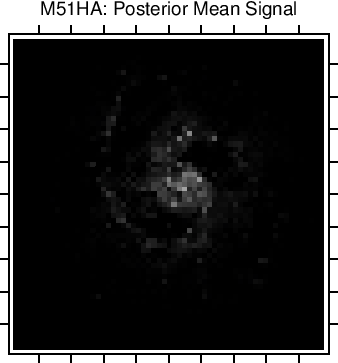}}
  {\includegraphics[type=png,ext=.png,read=.png,width=0.29\textwidth]{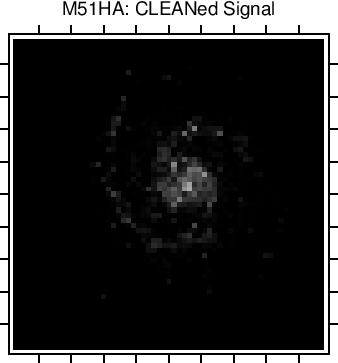}}
  
  {\includegraphics[type=png,ext=.png,read=.png,width=0.29\textwidth]{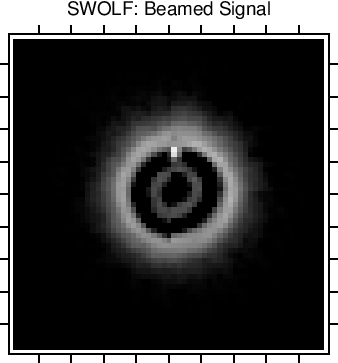}}
  {\includegraphics[type=png,ext=.png,read=.png,width=0.29\textwidth]{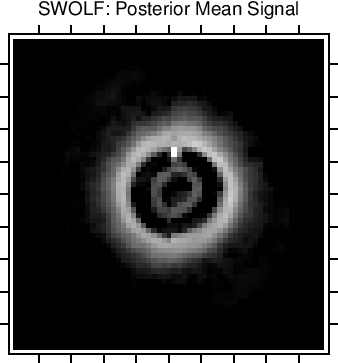}}
  {\includegraphics[type=png,ext=.png,read=.png,width=0.29\textwidth]{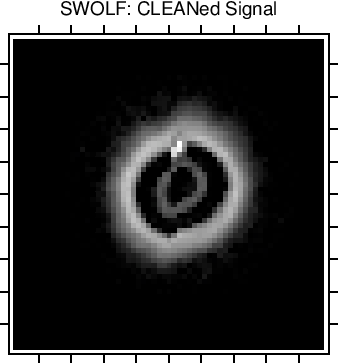}}
  
  {\includegraphics[type=png,ext=.png,read=.png,width=0.29\textwidth]{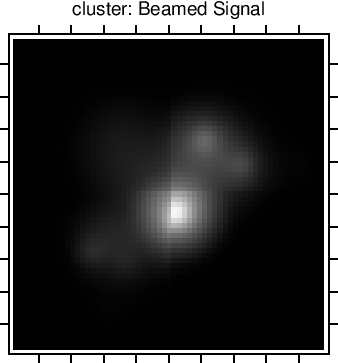}}
  {\includegraphics[type=png,ext=.png,read=.png,width=0.29\textwidth]{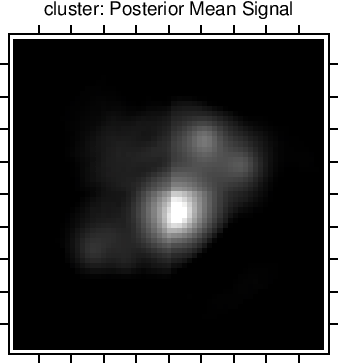}}
  {\includegraphics[type=png,ext=.png,read=.png,width=0.29\textwidth]{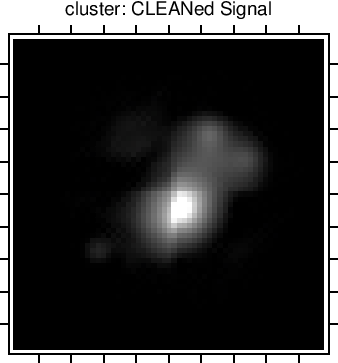}}
  
  {\includegraphics[type=png,ext=.png,read=.png,width=0.29\textwidth]{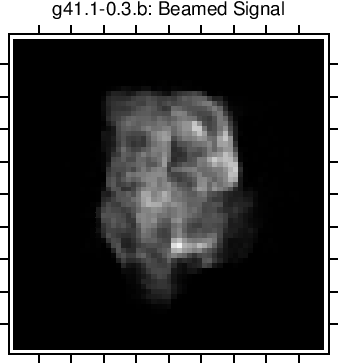}}
  {\includegraphics[type=png,ext=.png,read=.png,width=0.29\textwidth]{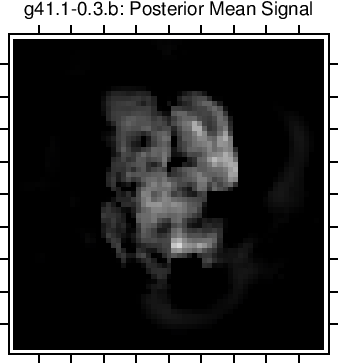}}
  {\includegraphics[type=png,ext=.png,read=.png,width=0.29\textwidth]{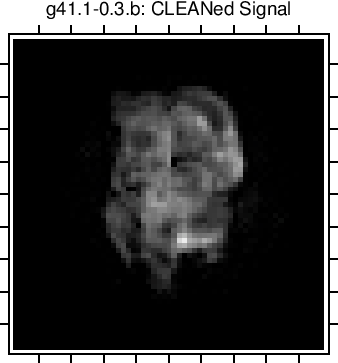}}

  \caption{
           Same as Figure~\ref{fig:gallery1} for the remaining four 
           test images.
          }
\label{fig:gallery2}
\end{figure*}

Figures~\ref{fig:gallery1} and~\ref{fig:gallery2} are galleries of 
the input images (without noise), our posterior mean reconstruction 
using Gibbs sampling, and a standard reconstruction using our 
implementation of {\tt CLEAN}. 
For all these images we have zoomed in to the inner 10 degrees 
where the primary beam selects the most prominent signal.
We have applied our variance-based acceptance mask 
for our Gibbs sampling reconstructions.
We see that Gibbs sampling is able 
to recover the complete range of source images to very high 
fidelity. We even faithfully recover glitches in the input 
image, such as the asymmetric protoplanetary disk in the 
\emph{SWOLF} image and discontinuous lobe structure in the 
\emph{3c288} image. This emphasizes the power of the Wiener Filter:
in regions of strong data support, our samples are driven 
to the data regardless of choice of prior. 
We also recover low-intensity portions of the image both in the 
center of the image and towards the edge of the primary beam, 
such as the regions between the spiral arms in the 
\emph{M51HA} image. However, there are other low-intensity regions, 
such as the lower portion of the mustache in the \emph{Einstein} 
image, where the signal is too low to distinguish it from the 
noise, and our mask rejects those pixels.

A comparison to the {\tt CLEAN} images especially highlights the abilities 
of our Gibbs sampling reconstruction technique. In all images, 
we recover a broader range of fluxes over a wider extent than 
{\tt CLEAN} (as in, for example, the \emph{cluster} reconstruction). 
Also, Gibbs sampling is able to recover portions of the 
image further into the edges of the primary beam, as can be easily 
seen in the \emph{3c288} and \emph{HCO+4-360} images.
Finally, in images with glitches, such as \emph{SWOLF}, {\tt CLEAN} tends to 
exaggerate the asymmetries. 

\begin{figure*}
  \centering 
  {\includegraphics[type=png,ext=.png,read=.png,width=0.24\textwidth]{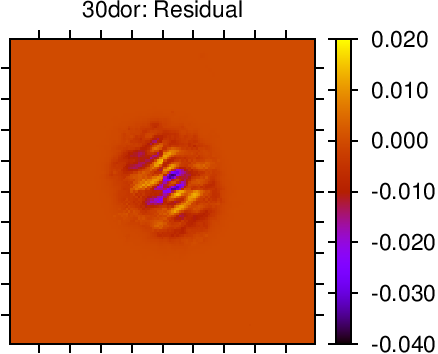}}
  {\includegraphics[type=png,ext=.png,read=.png,width=0.24\textwidth]{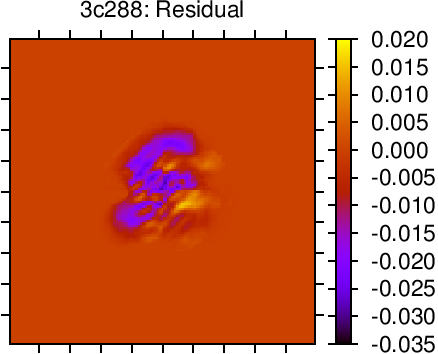}}
  {\includegraphics[type=png,ext=.png,read=.png,width=0.24\textwidth]{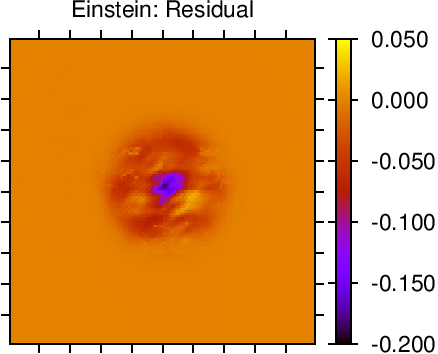}}
  {\includegraphics[type=png,ext=.png,read=.png,width=0.24\textwidth]{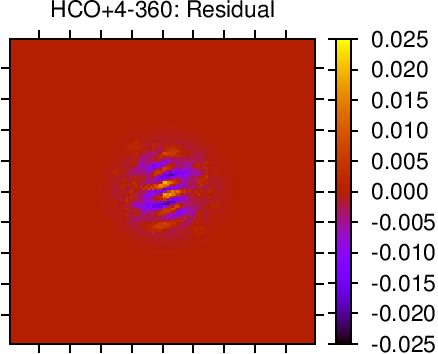}}
  {\includegraphics[type=png,ext=.png,read=.png,width=0.24\textwidth]{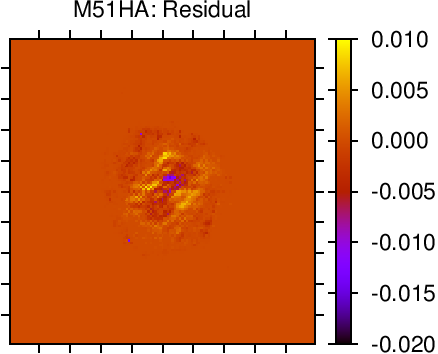}}
  {\includegraphics[type=png,ext=.png,read=.png,width=0.24\textwidth]{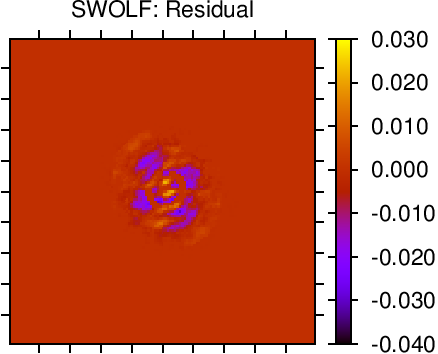}}
  {\includegraphics[type=png,ext=.png,read=.png,width=0.24\textwidth]{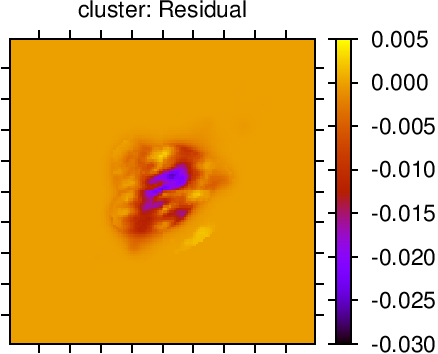}}
  {\includegraphics[type=png,ext=.png,read=.png,width=0.24\textwidth]{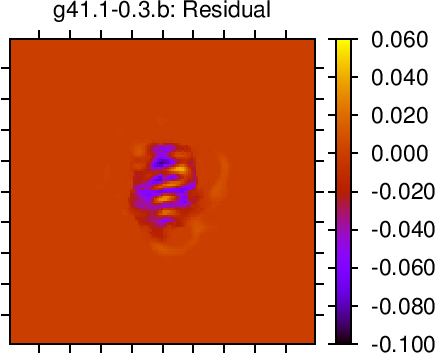}}
  \caption{
            Residuals (differences between reconstructed signal
            and beamed input signal $As$) for the Gibbs sampling algorithm
            for all test images. All images are 20 deg across.
            The color scale ranges from $0.0$ (black) to $1.0$ (white).
          }
\label{fig:residuals}
\end{figure*}

Figure~\ref{fig:residuals} shows the residuals (i.e., difference maps)
between the mean posterior signal generated by Gibbs sampling 
and the input test image. In all cases the variance-based acceptance 
mask has been applied. We see that the maximum residual occurs 
in the center of the \emph{Einstein} test image, where the Gibbs method
slightly underestimates the source intensity. This is probably 
due to the method attempting to average the 
very high-contrast features adjacent to the center.
The \emph{Einstein} image turns out to be the most difficult:
the residuals for all other images are typically an order of magnitude 
smaller. While the residuals are somewhat correlated 
with the distributions of the source image, like the \emph{Einstein} 
test image it appears that our method performs poorest 
in regions of high intensity contrast. However, these differences 
are very mild: typically on the order of 1\% of the input source 
intensity.

\begin{figure*}
  \centering 
  {\includegraphics[type=png,ext=.png,read=.png,width=0.24\textwidth]{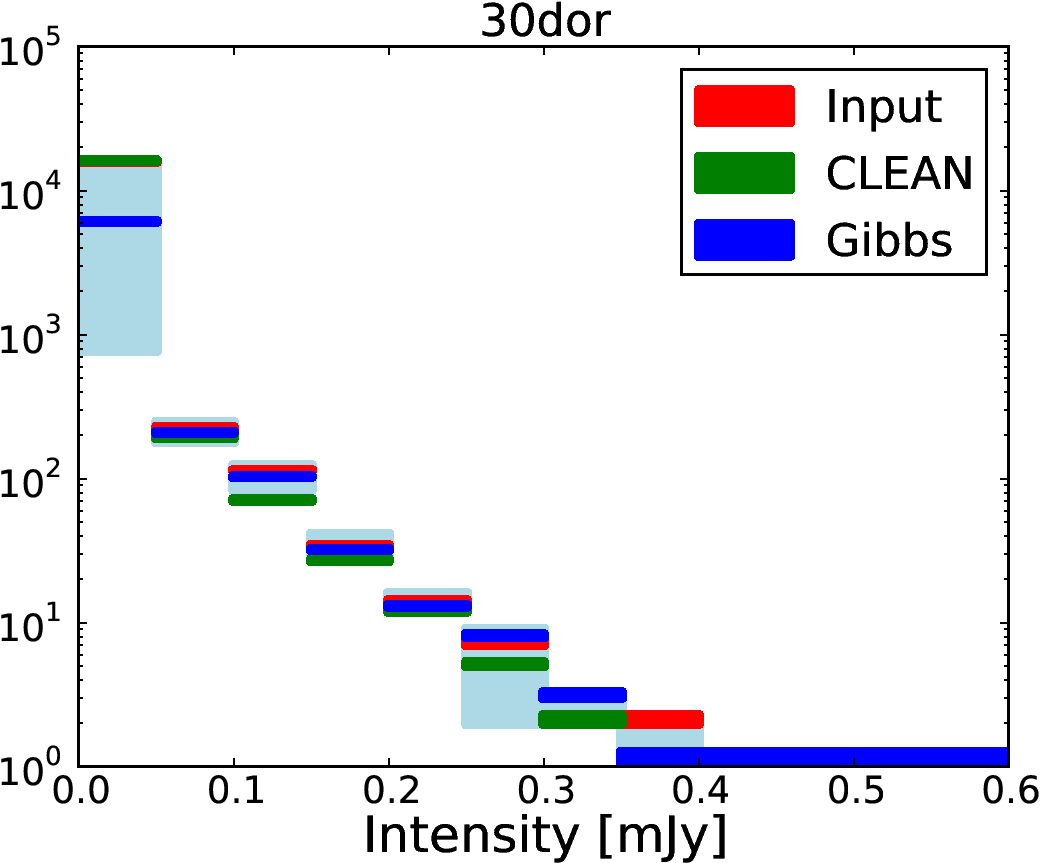}}
  {\includegraphics[type=png,ext=.png,read=.png,width=0.24\textwidth]{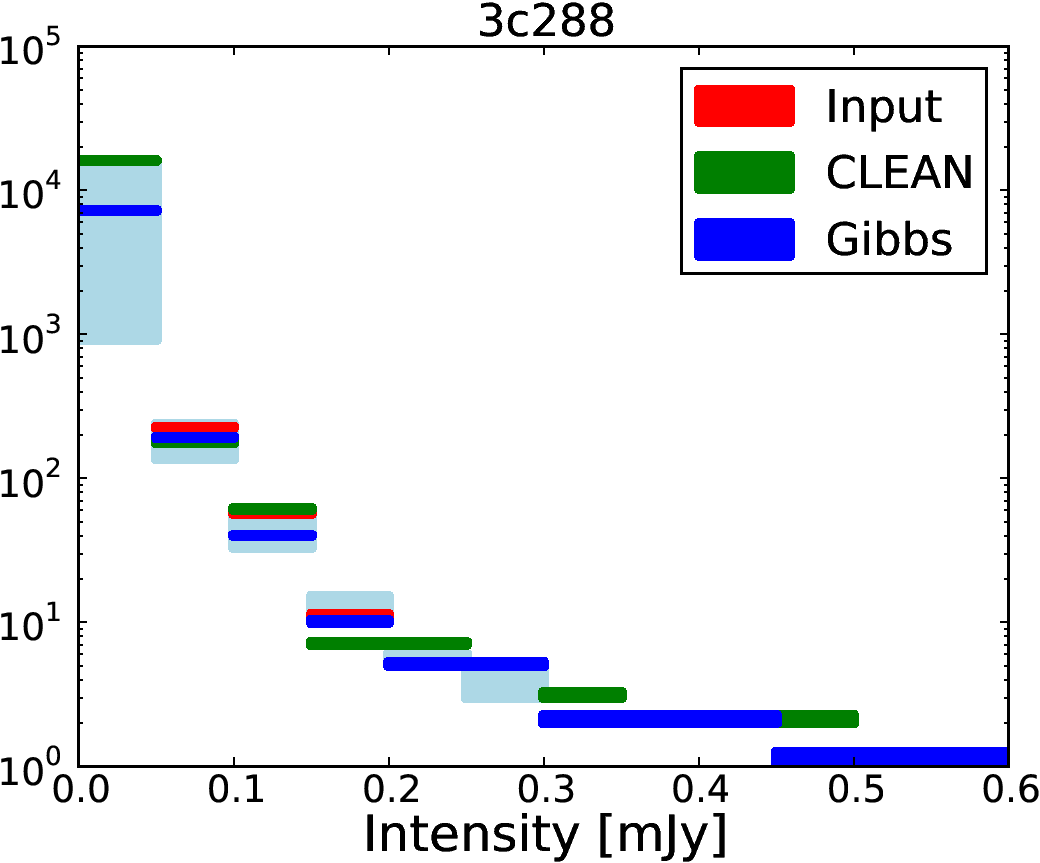}}
  {\includegraphics[type=png,ext=.png,read=.png,width=0.24\textwidth]{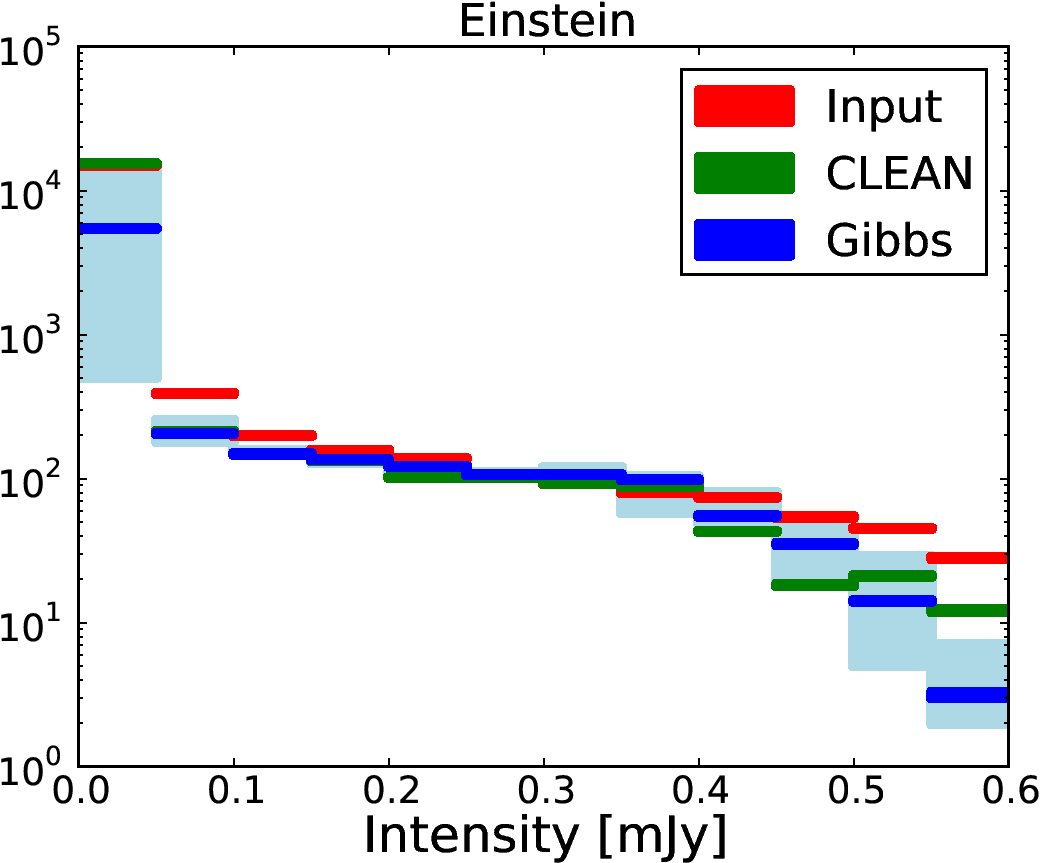}}
  {\includegraphics[type=png,ext=.png,read=.png,width=0.24\textwidth]{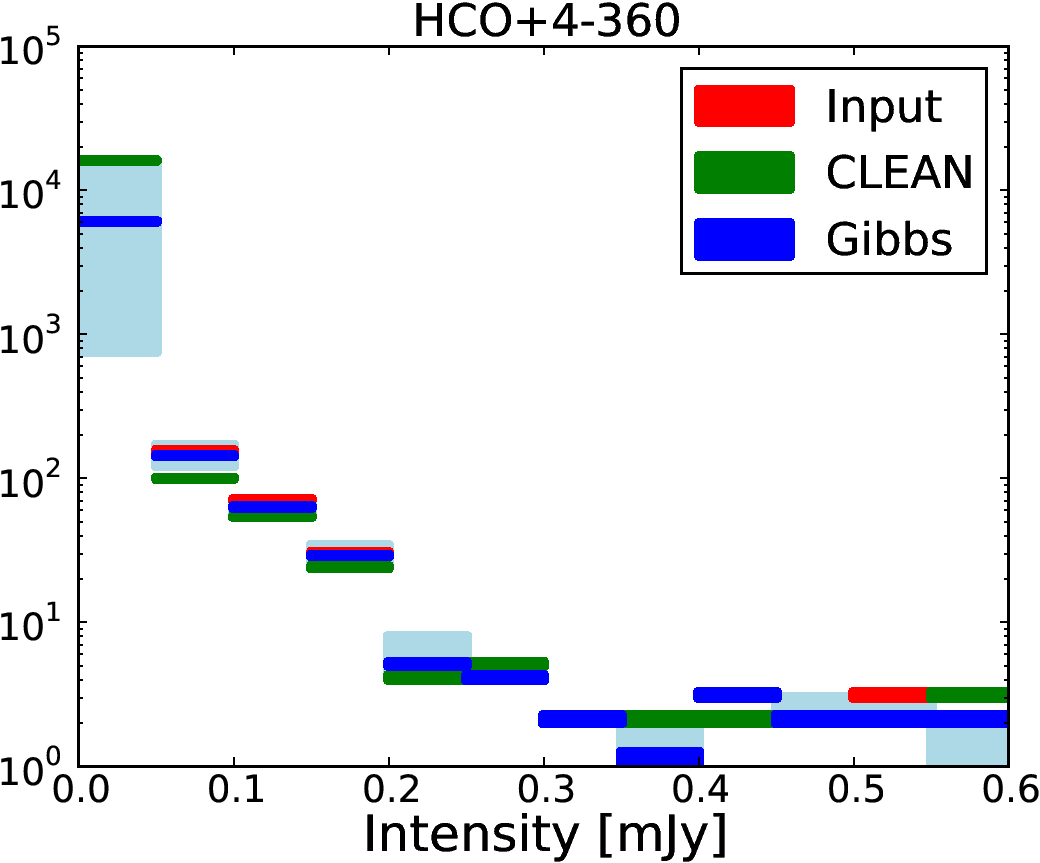}}
  {\includegraphics[type=png,ext=.png,read=.png,width=0.24\textwidth]{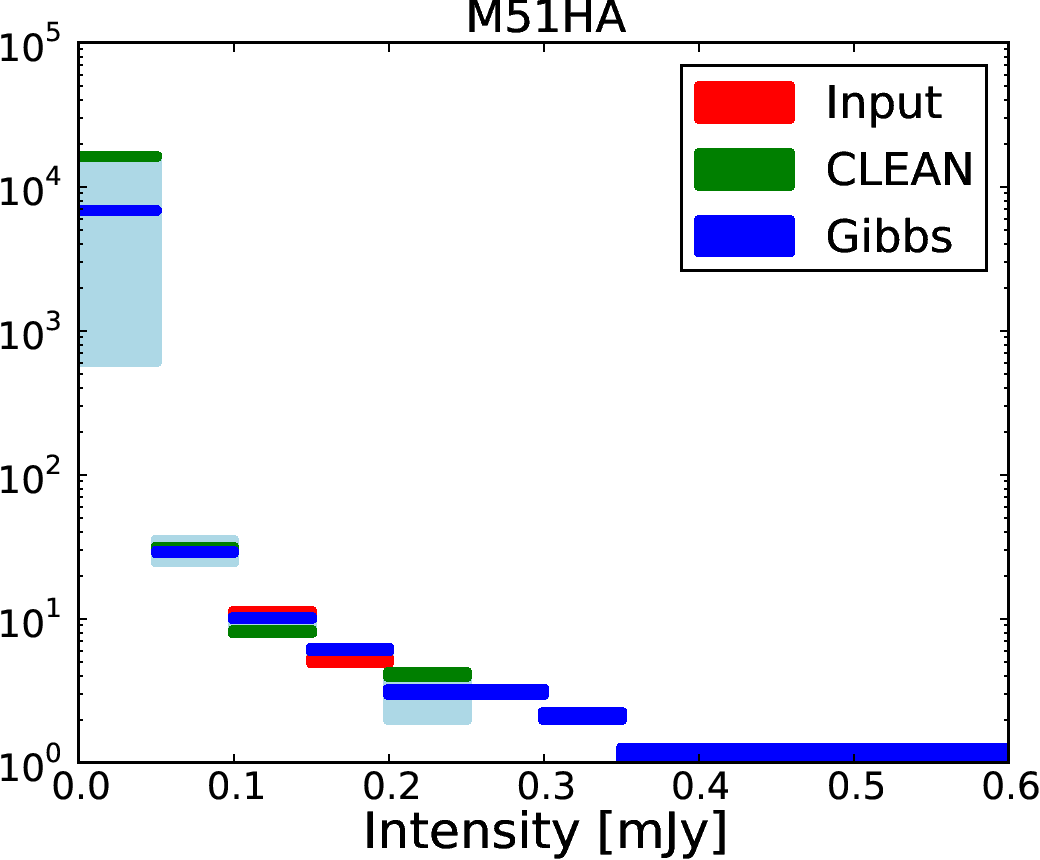}}
  {\includegraphics[type=png,ext=.png,read=.png,width=0.24\textwidth]{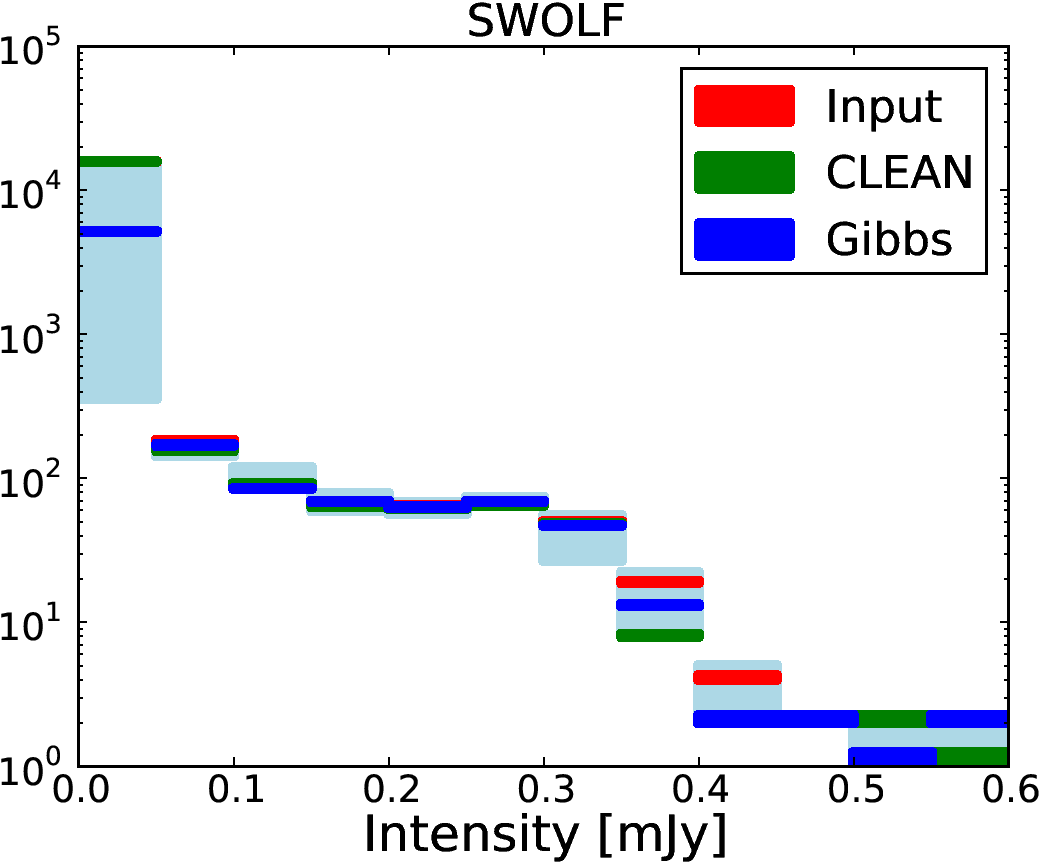}}
  {\includegraphics[type=png,ext=.png,read=.png,width=0.24\textwidth]{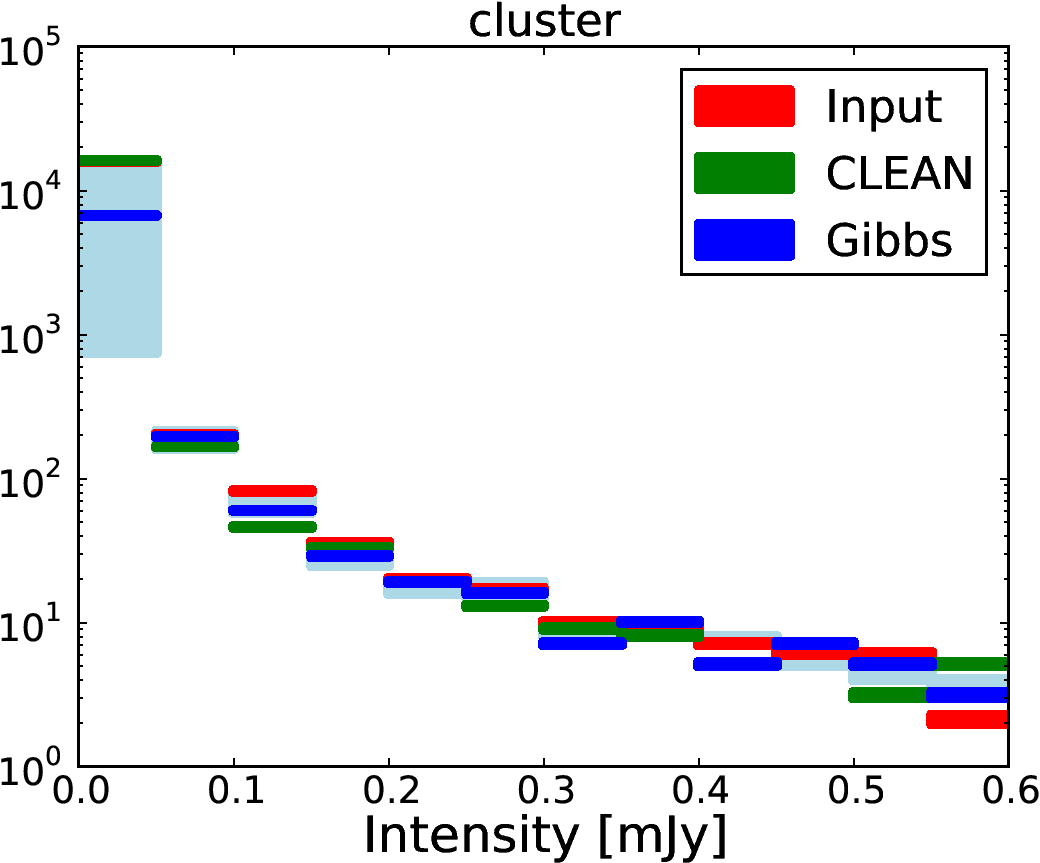}}
  {\includegraphics[type=png,ext=.png,read=.png,width=0.24\textwidth]{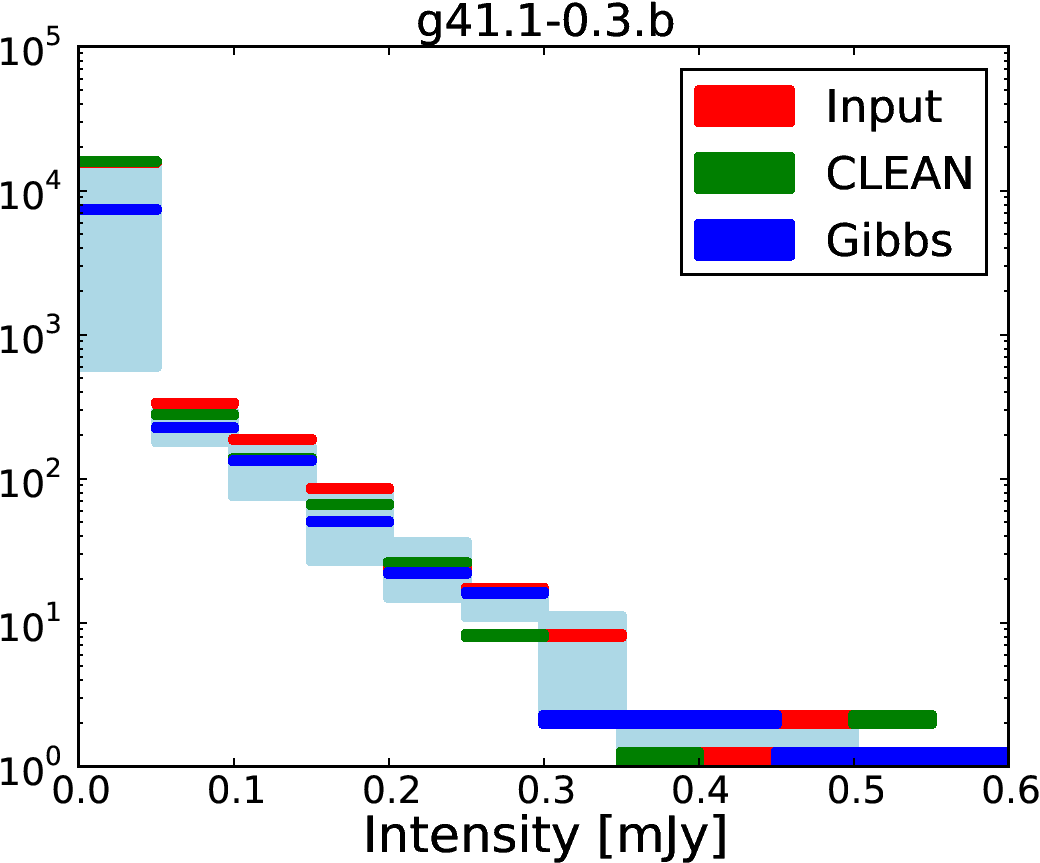}}
  \caption{
           Histograms of sky intensities in the beamed input signal $As$
           (red), {\tt CLEAN} reconstructed sky (green), and Gibbs sampling
           reconstructed sky (blue). Error bars (light blue) 
           on the Gibbs sampling
           intensities are 2$\sigma$ uncertainties calculated 
           from the posterior signal samples.
          }
\label{fig:histograms}
\end{figure*}

We may further quantify the differences between the input signals 
and their reconstructions with {\tt CLEAN} and our Gibbs sampling method 
by binning the intensities, as we do in Figure~\ref{fig:histograms}.
Bounding the histograms of the posterior mean are 
2$\sigma$ error bars measured from the variance in the generated samples.
This is another example of the kind of information unavailable 
in traditional reconstruction techniques. For almost all test 
images and intensity bins, our posterior mean reconstruction 
is within two standard deviations of the input signal. 
This is unsurprising: 
our Bayesian method automatically discovers the local 
variance because that variance is directly related to the relative
level of signal and noise in a given bin.
In a few bins, such as the high-intensity bin of the \emph{Einstein} 
test image, our method tends to underestimate the true 
intensity. This is due to the high pixel-by-pixel contrast 
in the central image region discussed above.

While the {\tt CLEAN} image intensities largely fall within the 
uncertainty ranges of the Gibbs reconstruction, there is a
systematic steepening of the distributions: {\tt CLEAN} tends to have too 
many low-intensity pixels and correspondingly too many 
high-intensity pixels. This validates the discussion above 
which noted that {\tt CLEAN} does not fully reproduce the observed 
range of input fluxes.

Finally, we may further simplify the comparison by reducing our 
measurement error to a single scalar. Two error metrics are 
commonly used: the root-mean-square (RMS) of the residual map, and 
the signal-to-noise ratio (SNR). This last quantity itself has 
several variations; we take that of~\citet{carrillo:sara}:
\begin{equation}
  {\rm SNR} \equiv 20 \log_{10} \frac{\sigma_s}{\sigma_{s-\hat{s}}},
\label{eq:snr}
\end{equation}
where $\sigma_x$ is the standard deviation of image x, 
with $s$ denoting the original image 
and $\hat{s}$ denoting the reconstructed image.
For both measures we first apply our variance-based acceptance 
mask before calculating the error metrics.

\begin{figure*}
  \centering 
  {\includegraphics[type=png,ext=.png,read=.png,width=0.48\textwidth]{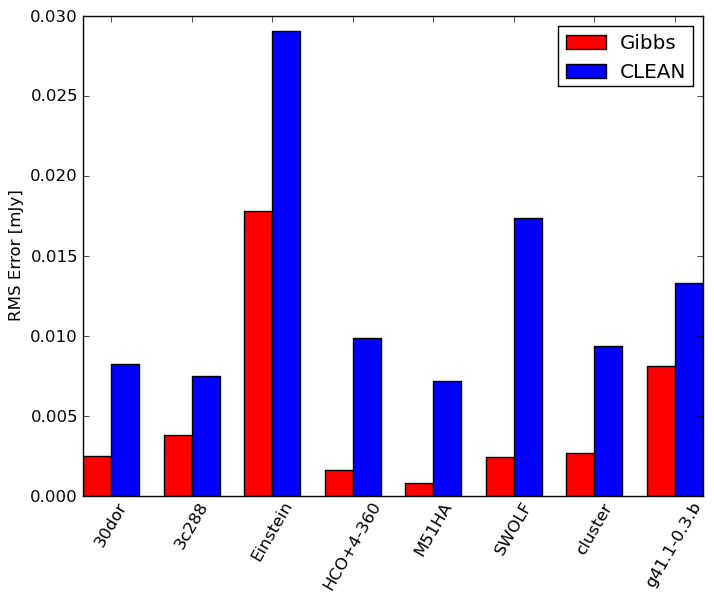}}
  {\includegraphics[type=png,ext=.png,read=.png,width=0.46\textwidth]{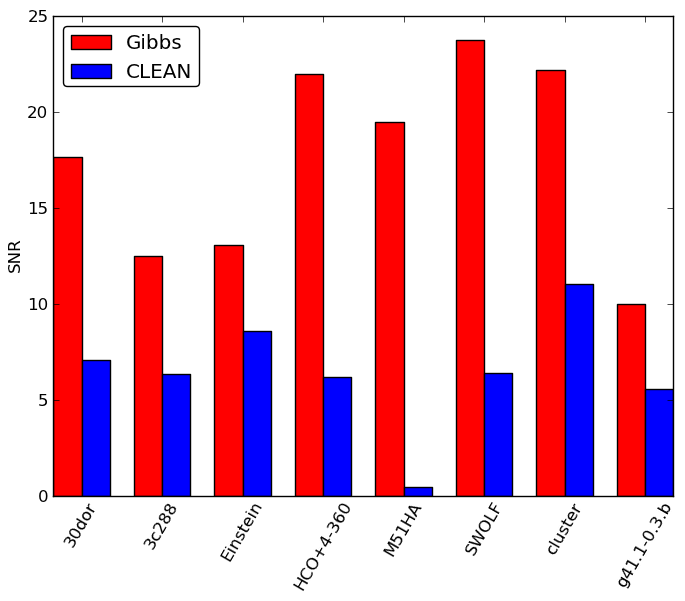}}
  \caption{
           Quantitative assessment of the errors. 
           We show RMS error between the beamed input signal $As$ 
           and the reconstructed images
           in the left plot.
           In the right plot we show the SNR as defined by
           Eq.~(\ref{eq:snr}).
           For each image we compare the Gibbs sampling reconstructed sky
           (red) to the {\tt CLEAN} reconstruction (blue).
           For the RMS (SNR) error measurement, smaller (larger) 
           values indicate better performance.
          }
\label{fig:error}
\end{figure*}

Figure~\ref{fig:error} shows the RMS and SNR for each of our 
test images for both Gibbs sampling and traditional {\tt CLEAN}.
As expected from our residuals, Gibbs sampling performed 
most poorly in terms of RMS with the \emph{Einstein} test image due to its 
highly complex structure. 
With the exception of that image and \emph{g41.1-0.3.b}, all RMS 
errors are below $4 \times 10^{-3}$. The SNR for all reconstructions 
with Gibbs sampling fall between 10 and 25.
In both RMS and SNR measures Gibbs sampling outperforms {\tt CLEAN} for 
all test images. The largest differences occur in 
images with high degrees of symmetry
(e.g., \emph{M51HA}, \emph{SWOLF}) since in these cases 
our isotropic Gaussian process prior performs best in marginal 
signal-to-noise portions of the image.

\section{Conclusions}
\label{sec:conclusion}

We have presented an innovative method for deconvolving radio 
interferometric images using the Bayesian method of Gibbs sampling. 
Our method naturally offers several advantages over traditional 
deconvolution approaches. It fully accounts for signal and 
noise mode coupling 
and incomplete $uv$-plane coverage in an automatic and 
well-motivated fashion, requires no fine-tuning or supervision 
as the method progresses, and provides an informative description 
of the uncertainty information in the signal reconstruction.
We choose an isotropic Gaussian process image prior, though 
we don't specify the signal covariance in advance.

We have tested our method with a realistic interferometric 
observing scenario of a wide variety of source images. 
These images represent typical targets, including protoplanetary 
disks and active galactic nuclei jet-and-lobe systems.
Note that our choice of prior is wrong for \emph{all test cases}, but the 
iterative application of the Wiener filter allows us to 
discover the source images within the noise and incomplete 
$uv$-plane coverage. 
We find that our method is quite robust: regardless of the 
structure of the source image we are able to 
recover the intensity distribution to a very high 
fidelity. Our method outperforms traditional point source-based 
{\tt CLEAN} in terms 
of intensity distributions, RMS error, and reconstruction 
signal-to-noise ratio. As expected given our isotropic Gaussian 
process prior, 
we perform best on images with large amount of symmetry, though 
the Wiener filter provides a route for reliable reconstructions of 
asymmetric images in regions of strong data support regardless 
of source structure.

Using our method we can also easily incorporate uncertainty information 
in the reconstructed image. 
As discussed in earlier works~\citep{Wandelt2004}, the error model 
generated from the isotropic Gaussian process prior tends to 
underestimate the true errors, but this situation is far preferable 
to no error model at all.
With our error model we can construct acceptance masks 
based on the local (i.e., pixel-by-pixel) signal-to-variance 
ratios. 
The desired threshold can be adjusted based on the 
desired level of confidence in the deconvolution. Since the sample
variance is directly tied to the relative level of noise 
by way of the Wiener Filter, our method naturally and 
self-consistently determines this mask.

While the test cases we have presented in this work 
use an interferometric setup with realistic noise levels, 
primary beam shape, and antenna array, 
they do represent a relatively simplistic observing scenario.
A fully implemented method would include simultaneous 
solutions of multiple frequencies, mosaicked images,
and wide bandwidth observations.
Also, we have specified an overall SNR of 10 for our fiducial 
observations, and we must examine the 
performance of this method in different regimes.

Additionally, future large-scale interferometers will deliver
incredibly high volumes of data. Image deconvolution from even 
a single pointing at a single frequency will tax most 
computing systems. 
To accommodate future data sets
we have implemented our algorithm using 
the MPI-parallelized {\tt PETSc}~\citep{petsc-web-page} toolkit, 
so our approach automatically 
grows with the size of the data without loss of scalability.
Algorithmically, the most expensive portion of our approach is the solution to
Equation~(\ref{eq:sky}), which scales 
as $\mathcal{O} (n_p \log n_p)$ in the ideal pregridded flat-sky approximation 
we have presented here. 

While we have not discussed foreground removal in this work, this method 
easily accommodates modeling in two fashions. First, partial signal or 
foreground information, if known in advance, can enter as an additional 
prior. Alternatively, foreground models can be added as an additional 
sampling step within the algorithm~\citep{Wandelt2004}. 
The resulting posterior mean signal will thus \emph{automatically} 
include marginalizations over the unconstrained parameters of 
the model. The same approach can be taken with the noise: 
if, for example, the noise spectrum is known but the absolute 
amplitude is not, we can sample over that amplitude. 
This flexibility, in addition to the other strengths 
discussed above, suggest that Gibbs sampling is a promising
response to the challenges of contemporary and future 
radio interferometric observations.

\section*{Acknowledgments}

P. M. Sutter and B. D. Wandelt acknowledge
support from NSF Grant AST-0908902. B. D. Wandelt also
acknowledges funding from an ANR Chaire d'Excellence,
the UPMC Chaire Internationale in Theoretical Cosmology, and NSF AST-0708849.
J.D. McEwen is supported in part by a Newton International Fellowship from the Royal Society and the British Academy. 
L. Zhang and P. Timbie acknowledge support from NSF
Grant AST-0908900. E. F. Bunn acknowledges support from NSF Grant AST-0908900.
G. S. Tucker and
A. Karakci acknowledge support from NSF Grant AST-
0908844.

\footnotesize{
  \bibliographystyle{mn2e}
  \bibliography{gibbsgas}
}

\end{document}